\documentclass{PoS}
\usepackage{amsmath}
\usepackage{amsthm}
\usepackage{graphicx}
\usepackage{caption}
\usepackage{subcaption}
\usepackage{multirow}
\allowdisplaybreaks
\def\Feynarts{{{\sc FeynArts}}}
\def\Feyncalc{{{\sc FeynCalc}}}

\newcommand{\unipd}{Dipartimento di Fisica ed Astronomia, Universit\`a
  di Padova, and INFN, Sezione di Padova, Via Marzolo 8, 35131 Padova, Italy}
\newcommand{\higgsct}{Higgs Centre for Theoretical Physics, School of Physics and Astronomy, The University of Edinburgh, Edinburgh EH9 3JZ, Scotland, UK}

\title{Adaptive Integrand Decomposition}

\ShortTitle{Adaptive Integrand Decomposition}

\author{\speaker{Pierpaolo Mastrolia}\\
        \unipd \\
        \email{ pierpaolo.mastrolia@pd.infn.it}}

\author{Tiziano Peraro \\
        \higgsct \\
        \email{ tiziano.peraro@ed.ac.uk}}

\author{ Amedeo Primo \\
        \unipd \\
        \email{ amedeo.primo@pd.infn.it}}

\author{William J. Torres Bobadilla\\
        \unipd \\
        \email{ william.torres@pd.infn.it}}

\abstract{We present a simplified variant of the integrand reduction
  algorithm for multiloop
  scattering amplitudes in $d = 4 - 2\epsilon$ dimensions, which exploits the decomposition of the
  integration momenta in parallel and orthogonal subspaces, $d=d_\parallel+d_\perp$, 
  where $d_\parallel$ is the dimension of the space spanned by the legs of the
  diagrams. We discuss the advantages of a lighter polynomial division
  algorithm and how the  orthogonality relations for Gegenbauer
  polynomilas can be suitably
  used for carrying out the integration of the irreducible monomials, which
  eliminates spurious integrals. 
  Applications to one- and two-loop integrals, for arbitrary
  kinematics, are discussed.
}

\FullConference{Loops and Legs in Quantum Field Theory\\
		24-29 April 2016\\
		Leipzig, Germany}

\begin{document}

\section{Introduction}
The decomposition of multiloop scattering amplitudes in terms of independent
functions, together with the subsequent determination of the
latter, is a viable alternative to
the direct integration which, for non-trivial processes,
may require the calculation of a prohibitively large 
 number of complicated Feynman integr{\it als}.

Decomposing multi-loop amplitudes in terms of independent integrals
can become problematic when the number of the scales of the diagrams
increases, due to the exchange or to the production of massive
particles, or when a large number of external particles are scattered, or
when the morphology of the contributing diagrams becomes involved. 
The integr{\it and} decomposition algorithm has the advantage of 
treating scattering amplitudes involving massive particles at the same
{\it price} of amplitudes for massless scattering. 
The output of the reduction procedure is the partial fractioning of the original integrand,
namely the determination of the remainders of the successive division
between the numerator and
(the partitions of the product of) the denominators. Upon inte gration, the partial
fraction formula correspond to rewrite the original amplitudes as a
combination of independent integrals.
However, the result of the integrand decomposition represents
an intermediate step towards the complete amplitude reduction. In
fact, additional relations among those integrals, like
integration-by-parts identities, can minimise the number of independent master
integrals (MIs) which can appear in the final formulas.

The integrand decomposition algorithm \cite{Ossola:2006us,Ossola:2007bb,Ellis:2007br,Ellis:2008ir,Ossola:2008xq,Mastrolia:2008jb,Mastrolia:2012bu} played a key role for 
the automation of one-loop corrections to high-multiplicity
scattering processes \cite{Ossolaproc}. The extension of this approach at two-loop and
beyond
\cite{Mastrolia:2011pr,Badger:2012dp,Zhang:2012ce,Mastrolia:2012an}
has been under intense investigation. 
The recent developments on the integrand side have been accompanied
by important developments for novel derivation of 
the integral relations needed to identify MIs \cite{Ita:2015tya,Larsen:2015ped,VonManteuffel:2014ixa,Kant:2013vta}, as well as by progress in
the ability of computing the latter analytically
\cite{Henn:2013pwa,Argeri:2014qva,Papadopoulos:2014lla} as well as 
numerically \cite{Borowka:2015mxa,Smirnov:2015mct}. This vivid research has been
largely due to the deeper understanding of the properties of
the integrands of Feynman graph, and of the refined algebraic and
differential calculus which control them.

In these proceedings, we summarise the results of ref. \cite{Mastrolia:2016dhn}.

\section{Parallel and orthogonal space for multiloop Feynman integrals}
In this contribution, we elaborate on a representation of dimensionally
regulated Feynman integrals where, for any given diagram, the number of space-time dimensions
$d \ ( = 4 - 2 \epsilon)$ is split into
{\it parallel} (or longitudinal) and {\it orthogonal} (or transverse)
dimensions, as $d=d_{\parallel}+ d_{\perp}$ \ 
\cite{Collins:105730,Kreimer:1991wj,Kreimer:1992zv,Czarnecki:1994td,Frink:1996ya,kreimer:1996qy}. 
Accordingly, the parallel space is spanned by the independent four-dimensional external momenta of the
diagram, namely $d_{\parallel} = n - 1$, where $n$ is the number of legs, whereas
the transverse space is spanned by the complementary orthogonal directions.
For diagrams with a number of legs 
$n \ge 5$, the 
orthogonal space embeds the $-2 \epsilon$ regulating dimensions, $d_\perp = - 2 \epsilon$,
while, for diagrams with $n \le 4$,
the orthogonal space is larger and it embeds, beside the regulating dimensions, also  
 the four-dimensional complement of the parallel space, namely  
 $d_\perp = (5-n) - 2 \epsilon$. In this sense, the decomposition of
 the space-time dimensions in parallel and orthogonal directions can
 be considered as {\it adaptive}, since it depends on the number
 of legs of the individual diagram.\\

A generic $\ell$-loop Feynman integral with $n$ external legs in a
$d$-dimensional Euclidean space can be written as
\begin{align}
  I_{n}^{d\,(\ell)}[\mathcal{N}]=\int\left( \prod_{i=1}^{\ell}\frac{d^dq_{i}}{\pi^{d/2}}\right)\frac{\mathcal{N}(q_{i})}{\prod_{j}D_{j}(q_{i})}.
  \label{eq:Ilk+1}
\end{align}
In the previous equation $\mathcal{N}(q_{i})$ is an arbitrary tensor
numerator and the denominators $D_j(q_i)$ are quadratic in the loop
momenta and can be written as
\begin{align}
  &D_j=l_j^2+m_j^2, \quad\text{with}\quad l_j^{\alpha}=\sum_{i}\alpha_{ij}q_i^{\alpha}+\sum_{i}\beta_{ij}p_i^{\alpha},
    \label{eq:dens2l}
\end{align}
where $\{p_1,\,\dots,\,p_{n-1}\}$ is the set of independent external
momenta and $\alpha,\beta \in\{0,\pm 1\}$.

When dealing with regularisation schemes where the external kinematics
is kept in four dimensions, the $d$-dimensional loop momenta are often
split into a four-dimensional part and a $(-2\epsilon)$-dimensional
one,
		\begin{align}
		q^{\alpha}_{i}=&q_{[4]\,i}^{\alpha}+\mu_i^{\alpha},\qquad q_i\cdot q_{j}=q_{[4]\, i}\cdot q_{[4]\, j}+\mu_{ij},\quad (\mu_{ij}=\mu_{i}\cdot\mu_{j}),
		\label{4mu}
		\end{align}
        and the denominators read
		\begin{align}
		&D_i=l_{i[4]}^2+\sum_{j,k}\alpha_{ij}\alpha_{ik}\,\mu_{jk}+m_i^2,\quad\text{with}\quad l_{i[4]}^{\alpha}=\sum_{j}\alpha_{ij}q_{i[4]}^{\alpha}+\sum_{j}\beta_{ij}p_j^{\alpha}.
		\label{eq:densmu}
		\end{align}
		Therefore both the numerator in \eqref{eq:Ilk+1} and the
        denominators become polynomials in ${\ell(\ell+9)}/{2}$
        variables, namely the $(-2\epsilon)$-dimensional scalar
        products $\mu_{ij}$ and the components of
        $q_{i[4]}^{\alpha}$ with respect to a four-dimensional basis
        of vectors $\{e_i^{\alpha}\}$,
      $q_{[4]\, i}^{\alpha}=\sum_{j=1}^{4}x_{ji}e_j^{\alpha}$.
       Thus, in $d=4-2\epsilon$, we can write
		\begin{align}
		I_{n}^{d\,(\ell)}[\mathcal{N}]=\Omega^{(l)}_d\int\prod_{i=1}^{\ell} d^4q_{[4]\,i}\int \prod_{1\leq i\leq j\leq \ell} d\mu_{ij}\left[G(\mu_{ij})\right]^{\frac{d-5-\ell}{2}} \frac{\mathcal{N}(q_{[4]\,i},\mu_{ij})}{\prod_{m}D_{m}(q_{[4]\,i},\mu_{ij})},
		\label{eq:oldpar}
		\end{align}
		where $G(\mu_{ij})=\text{det}[(\mu_i\cdot\mu_j)]$ is the Gram
        determinant and the prefactor $\Omega^{(\ell)}_d$ is the
        result of the angular integration over the angular directions.

        For a number of external legs $n\leq 4$, the external momenta
        define a $d_{\parallel}$-dimensional subspace with
        $d_{\parallel}=n-1$.  In these cases one can parametrise the
        integral \eqref{eq:Ilk+1} in such a way that the number of
        variables appearing in the denominators is reduced to
        ${\ell(\ell+2d_{\parallel}+1)}/{2}$.  The numerator will
        instead still have a polynomial dependence over the remaining
        $\ell(4-d_{\parallel})$ variables.  These can however be
        integrated out via a straightforward expansion of the
        numerator in terms of orthogonal polynomials.  More in detail,
        if $d_{\parallel}\leq 3$, one can
        choose $4-d_{\parallel}$ of the vectors in the basis $\{e_i^{\alpha}\}$ to lie into the
        subspace orthogonal to the external kinematics, \textit{i.e.} 
        $e_i\cdot p_j=0\: (i> d_{\parallel},\,\forall j)$, and 
        $e_i\cdot e_j =\delta_{ij} \;(i,j> d_{\parallel})$.
		 In this way, loop momenta in $d=d_{\parallel}+d_{\perp}$  read
		 \begin{align}
		 q_{i}^{\alpha}=q_{\parallel\, i}^{\alpha}+\lambda^{\alpha}_{i},
		 \label{eq:newdeco}
		 \end{align}
		 \begin{align}
		 q_{\parallel\, i}^{\alpha}=\sum_{j=1}^{d_{\parallel}}x_{ji}e_{j}^{\alpha},\qquad
		 \lambda^{\alpha}_i=\sum_{j=d_{\parallel}+1}^4x_{ji}e_j^{\alpha}+\mu_i^{\alpha},
		 \end{align}
		 where $q_{\parallel\, i}$ is a vector of the $d_{\parallel}$-dimensional space spanned by the external momenta,
		and $\lambda_{i}$ belongs the $d_{\perp}$-dimensional orthogonal subspace. In this parametrisation, all denominators become independent of the transverse components of the loop momenta,
		 \begin{align}
		 &D_i=l_{\parallel\, i}^2+\sum_{j,l}\alpha_{ij}\alpha_{il}\,\lambda_{jl}+m_i^2,\quad
		 &l_{\parallel\, i}^{\alpha}=\sum_{j}\alpha_{ij}q_{\parallel\, i}^{\alpha}+\sum_{j}\beta_{ij}p_j^{\alpha},
		 \qquad \lambda_{ij}=\sum_{l=d_{\parallel}+1}^4x_{li}x_{lj}+\mu_{ij},
		 \label{eq:dens2l2}
		 \end{align}
		 and they depend on a reduced set of
         ${\ell(\ell+2d_{\parallel}+1)}/{2}$ variables, corresponding
         to the $\ell d_{\parallel}$ components of
         $q_{\parallel\, i}^{\alpha}$ and the $\ell(\ell+1)/2$ scalar
         products $\lambda_{ij}$. 
        In $d=d_{\parallel}+d_{\perp}$, the integral \eqref{eq:Ilk+1} can thus be rewritten as
		\begin{align}
		I_{n}^{d\,(\ell)}[\mathcal{N}]=\Omega^{(\ell)}_d\!\int\prod_{i=1}^{\ell} d^{n-1}q_{\parallel \, i}\int\! d^{\frac{\ell(\ell+1)}{2}} \boldsymbol{\Lambda}\int\! d^{(4-d_{\parallel})\ell}\boldsymbol{\Theta}_{\perp}\frac{\mathcal{N}(q_{i \,\parallel},\boldsymbol{\Lambda},\boldsymbol{\Theta}_{\perp})}{\prod_{j}D_{j}(q_{ \parallel\, i},\boldsymbol{\Lambda})},
		\label{eq:lambth}
		\end{align}
		where
		\begin{subequations}
		\begin{align}
		\label{eq:lambdaShout}
		\int d^{\frac{\ell(\ell+1)}{2}} \boldsymbol{\Lambda}
		=\int \prod_{1\leq i\leq j}d\lambda_{ij}\left[G(\lambda_{ij})\right]^{\frac{d_{\perp}-1-\ell}{2}}
		\end{align}
		\end{subequations}
		defines the integral over the norm of the  transverse vectors
        $\lambda_{i}^{\alpha}$ and their relative orientations and
        $\boldsymbol{\Theta}_{\perp}$ parametrises the integral over
        the components of $\lambda_{i}^{\alpha}$ lying in the
        four-dimensional space.
	
	    Remarkably, eq.~\eqref{eq:newdeco}
            allows to express a subset of components of
            $q_{\parallel i}^\alpha$ and $\lambda_{ij}$ as
            combinations of loop denominators by solving linear
            relations. Therefore, one can always build differences of
            denominators which are linear in the loop momenta and
            independent of $\lambda_{ij}$, while the relation between
            $\lambda_{ij}$ and the denominators is always linear by
            definition, as it can be seen from eq.~\eqref{eq:dens2l2}. 

Since the denominators do not depend on the $\boldsymbol{\Theta}_{\perp}$-components,
their integration can be easily performed. In fact, the integration
over $\boldsymbol{\Theta}_{\perp}$ amounts to a
product of factorised, univariate integrations of polynomial integrands, each of the type
		\begin{align}
		\int_{-1}^{1}\!d\!\cos\theta_{ij} (\sin\theta_{ij})^{\alpha}(\cos\theta_{ij})^{\beta}.
		\label{eq:angularint}
		\end{align}
		The values of $\alpha$ and $\beta$ depend on the
                specific expression of the numerator. Nevertheless,
                these  integrals can be computed once and for all up
                to the desired rank and then re-used in every 
                calculation, when occurring. These integrals can be
                evaluated by performing the Passarino-Veltman tensor
                reduction in the orthogonal space. Alternatively, they
                can be evaluated by exploiting the properties of
                of \textit{Gegenbauer polynomials}
                $C^{(\alpha)}_{n}(\cos\theta)$, a particular class of
                orthogonal polynomials over the interval $[-1,1]$,
                which obey the orthogonality relation
			\begin{align}
			\int_{-1}^{1}\!d\!\cos\theta (\sin\theta)^{2\alpha-1}C^{(\alpha)}_{n}(\cos\theta)C^{(\alpha)}_{m}(\cos\theta)=\delta_{mn}\frac{2^{1-2\alpha}\pi\Gamma(n+2\alpha)}{n!(n+\alpha)\Gamma^2(\alpha)}.
			\label{eq:ortcos}
			\end{align}

We observe that there are special classes of multiloop integrals, associated to factorised and ladder topologies, whose denominators are independent of a certain number of  transverse orientations $\lambda_{ij}$. For these cases, the Gegenbauer integration can be applied, besides to all $\boldsymbol{\Theta}_{\perp}$, also to  the $\lambda_{ij}$ which do not appear in the denominators.

\section{Adaptive Integrand Decomposition }
	\subsection{Integrand recurrence relation}\label{sec:integrrecrel}
	In the framework of the integrand reduction method \cite{Ossola:2006us,Ellis:2007br, Mastrolia:2011pr, Badger:2012dp, Zhang:2012ce, Mastrolia:2013kca}, the computation of dimensionally regulated $\ell$-loop integrals
	\begin{align}
	I^{d\,(\ell)}_{i_1\dots i_r}=\int \prod_{j=1}^{\ell}\frac{d^dq_{j}}{\pi^{d/2}}\,\frac{\mathcal{N}_{i_1\dots i_r}(q_j)}{D_{i_1}(q_j)\cdots D_{i_r}(q_j)}
	\end{align}
	 is rephrased in terms of the reconstruction of the integrand function as a sum of integrands with irreducible numerators (or \textit{residues}) and a subset of denominators $D_{i_k}$,
	 \begin{align}
	 \mathcal{I}_{i_1\dots i_r}(q_j)\equiv\frac{\mathcal{N}_{i_1\dots i_r}(q_j)}{D_{i_1}(q_j)\cdots D_{i_r}(q_j)}=\sum_{k=0}^{r}\sum_{\{i_1\cdots i_k\}}\frac{\Delta_{j_1\cdots j_k}(q_j)}{D_{j_1}(q_j)\cdots D_{j_{k}}(q_j)}.
	 \label{eq:intdec}
	 \end{align}
	For an integral with an arbitrary number $n$ of external legs, the integrand decomposition formula \eqref{eq:intdec} can be obtained by observing that both numerator and denominators are polynomials in the components of the loop momenta with respect to some basis, which we collectively label as $\mathbf{z}=\{z_1,\dots,z_{\frac{\ell(\ell+9)}{2}}\}$. Thus, we can fix a monomial ordering and  build a Gr\"{o}ebner basis $\mathcal{G}_{i_1\cdots i_r}(\mathbf{z})$ of the ideal $\mathcal{J}_{i_1\cdots i_r}$ generated by the set of denominators,
	\begin{align}
	\mathcal{J}_{i_1\cdot\cdot \cdot i_r}\equiv \bigg\{\sum_{k=1}^{r}h_k(\mathbf{z})D_{i_k}(\mathbf{z})\,:\,h_k(\mathbf{z})\in P[\mathbf{z}]\bigg\},
	\end{align}
	being $P[\mathbf{z}]$ the ring of polynomials in $\mathbf{z}$. By performing the polynomial division of $\mathcal{N}_{i_1\cdots i_r}(\mathbf{z})$ modulo $\mathcal{G}_{i_1\cdot\cdot\cdot i_r}(\mathbf{z})$,
	\begin{align}
	\mathcal{N}_{i_1\cdots i_r}(\mathbf{z})=\sum_{k=1}^{r}\mathcal{N}_{i_1\cdots i_{k-1}i_{k+1}\cdots i_{r}}(\mathbf{z})D_{i_k}(\mathbf{z})+\Delta_{i_1\cdots i_r}(\mathbf{z})
	\label{eq:q+r}
	\end{align}
	we obtain the recurrence relation
	\begin{align}
	\mathcal{I}_{i_1\cdots i_r}=\sum_{k=1}^{r}\mathcal{I}_{i_1\cdots i_{k-1}i_{k+1}\cdots i_{r}}+\frac{\Delta_{i_1\cdots i_r}(\mathbf{z})}{D_{i_1}(\mathbf{z})\,\cdots D_{i_n}(\mathbf{z})\,},
	\label{eq:rec}
	\end{align}
	whose iterative application to the integrands corresponding to subtopologies with fewer loop propagators yields to the complete decomposition \eqref{eq:intdec}.\\
	
   Depending on the choice of variables $\mathbf{z}$ and the monomial order, the picture presented in this section can significantly simplify.  A particular convenient choice of variables turns out to be the one presented in eq.~\eqref{eq:newdeco}.  Indeed, as already observed, we can always express a subset of the components of $q_{\parallel i}^\alpha$ and $\lambda_{ij}$ as a combination of denominators by solving linear relations.  This set of relations is in turn equivalent to the definition of the denominators themselves.  This implies that if we choose the lexicographic monomial order with $\lambda_{ij}\prec x_{kl}$ for $k\leq d_{\parallel}$, the polynomials in the Gr\"obner bases are linear in the $\lambda_{ij}$ and the reducible components of $q_{\parallel i}^\alpha$.  The polynomial division can thus equivalently be performed by applying the aforementioned set of linear relations without explicitly computing the corresponding Gr\"obner basis.

   \subsection{Divide, integrate and divide}
    When dealing with an integral with $n\leq 4$ external legs, we can use the $d=d_{\parallel}+d_{\perp}$ parametrisation which removes the dependence of the denominators on the transverse components of the loop momenta. Thus, if we indicate with $\mathbf{z}$ the full set of $\ell(\ell+9)/2$ variables
	\begin{align}
	\mathbf{z}=&\{\mathbf x_{\parallel\,i},\mathbf x_{\perp\,i},\lambda_{ij}\},\quad i,j=1,\dots\ell,
	\end{align}
	where $\mathbf{x}_{\parallel\,i}$($\mathbf{x}_{\perp\,i}$) are the components of the loop momenta parallel(orthogonal) to the external kinematics, the denominators are reduced to polynomials in the subset of variables
	\begin{align}
	\boldsymbol{\tau}=&\{\mathbf{x}_{\parallel},\lambda_{ij}\},\quad \boldsymbol{\tau}\subset \mathbf{z},
	\end{align}
	so that the general $r$ denominators integrand has the form
	\begin{align}
	 \mathcal{I}_{i_1\dots i_r}(\boldsymbol{\tau},\mathbf{x}_{\perp})\equiv\frac{\mathcal{N}_{i_1\dots i_r}(\boldsymbol{\tau},\mathbf{x}_{\perp})}{D_{i_1}(\boldsymbol{\tau})\cdots D_{i_r} (\boldsymbol{\tau})}.
	\end{align}
	This observation suggests an \textit{adaptive} version of the integrand decomposition algorithm, where the polynomial division is simplified by working on the reduced set of variables $\boldsymbol{\tau}$ and the expansion of the residues in terms of Gegenbauer polynomials allows the systematic identification of spurious terms. The algorithm is organised in three steps:
	\begin{enumerate}
		\item \textbf{Divide:} we adopt lexicographic ordering $\lambda_{ij}\prec \mathbf x_{\parallel}$ for the $\boldsymbol{\tau}$  variables and we divide the numerator $\mathcal{N}_{i_1\dots i_r}(\boldsymbol{\tau},\mathbf{x}_{\perp})$ modulo the Gr\"{o}ebner basis $\mathcal{G}_{i_1\cdots i_r}(\boldsymbol{\tau})$ of the ideal $\mathcal{J}_{i_1\cdots i_r}(\boldsymbol{\tau})$ generated by the denominators,
		\begin{align}
		\mathcal{N}_{i_1\dots i_r}(\boldsymbol{\tau},\mathbf{x}_{\perp})=\sum_{k=1}^{r}\mathcal{N}_{i_1\dots i_{k-1}i_{k+1}\dots i_r}(\boldsymbol{\tau},\mathbf{x}_{\perp})D_{i_k}(\boldsymbol{\tau})+\Delta_{i_1\dots i_{r}}(\mathbf x_{\parallel},\mathbf x_{\perp}).
		\end{align}
		As a consequence of the specific monomial ordering, the residue $\Delta_{i_1\dots i_{r}}$ can depend on the transverse components $\mathbf x_{\perp\,i}$, which are left untouched by the polynomial division, as well as on $\mathbf x_{\parallel\,i}$ but not on $\lambda_{ij}$ that are expressed in terms of denominators and irreducible physical scalar products. Conversely, the quotient, from which the numerators corresponding to the subdiagrams to be further divided are obtained, still depends on the full set of loop variables.  As we explained at the end of sec.~\ref{sec:integrrecrel}, the Gr\"obner basis does not need to be explicitly computed, since, with the choice of variables and the ordering described here, the division is equivalent to applying the set of linear relations described above.
		\item\textbf{Integrate:} by writing the contribution of the residue $\Delta_{i_1\dots i_{r}}$ to the integral in the $d=d_{\parallel}+d_{\perp}$ parametrisation, we can integrate over transverse directions through the expansion of $\Delta_{i_1\dots i_{r}}$ in terms of Gegenbauer polynomials, which sets to zero spurious terms and reduce all non-vanishing contributions to monomials in $\lambda_{ij}$.
		It should be noted that, due to the angular prefactors produced by the integration of the transverse directions, the integrated residue
		\begin{align}
		\Delta^{\text{int}}_{i_1\dots i_{r}}(\boldsymbol{\tau})=\int\! d^{(4-d_{\parallel})\ell}\boldsymbol{\Theta}_{\perp}\Delta_{i_1\dots i_{r}}(\boldsymbol{\tau},\boldsymbol{\Theta}_{\perp})
		\end{align}
		is, in general, a polynomial in $\boldsymbol{\tau}$ whose coefficients depend explicitly on the space-time dimension $d$. The full set of $\Delta^{\text{int}}_{i_1\dots i_{r}}(\boldsymbol{\tau})$, obtained by iterating on each subdiagram numerator the polynomial division and the integration over the transverse space, provides already a spurious term-free representation of the integrand \eqref{eq:intdec}.
		\item\textbf{Divide:} however, since $\Delta^{\text{int}}_{i_1\dots i_{r}}(\boldsymbol{\tau})$ depends on the same variables as the denominators $D_{i_k}(\boldsymbol{\tau})$, we can perform a further division modulo the Gr\"{o}ebner basis $\mathcal{G}_{i_1\cdots i_r}(\boldsymbol{\tau})$ and get
		\begin{align}
		\Delta_{i_1\dots i_r}^{\text{int}}(\boldsymbol{\tau})=\sum_{k=1}^{r}\mathcal{N}^{\text{int}}_{i_1\dots i_{k-1}i_{k+1}\dots i_r}(\boldsymbol{\tau})D_{i_k}(\boldsymbol{\tau})+\Delta^{\prime}_{i_1\dots i_r}(\mathbf x_{\parallel}),
		\end{align}
		where, due to the choice of lexicographic ordering, the new residue $\Delta^{\prime}_{i_1\dots i_r}(\mathbf x_{\parallel})$ can only depend on $\mathbf x_{\parallel}$. Therefore, this additional polynomial division allows us to obtain an integrand decomposition formula \eqref{eq:intdec}, where all irreducible numerators are function of the components of the loop momenta parallel to the external kinematics.  As in the previous case, the division modulo Gr\"obner can equivalently be implemented via a set of linear relations.
	\end{enumerate}
	The interpretation of the monomials appearing in the residues $\Delta^{\prime}_{i_1\cdots i_r}(\mathbf{x}_{\parallel})$ in terms of a basis of tensor integrals can be additionally simplified by making use of the Gram determinant $G[\lambda_{ij}]$ (or $G[\mu_{ij}]$ for cases with more than four external legs, where $\mathbf{x}_\parallel\equiv \mathbf{x}$). In fact, as it can be easily understood from \eqref{eq:oldpar} and \eqref{eq:lambdaShout}, $G[\mu_{ij}]$ and  $G[\lambda_{ij}]$ can be interpreted as operators that, when acting on an arbitrary numerator of a $d$-dimensional integral, produce a dimensional shift $d\to d+2$. Therefore, Gram determinants can be used in order to trade some of the $d$-dimensional tensor integrals originating from the residues with lower rank integrals in higher dimensions.
	\subsection{Integrate and divide}
	In the three-step algorithm {\it divide-integrate-divide}, outlined in
	the previous section, the integration over the transverse angles is
	performed after the integrand reduction, namely after determining the
	residues. This first option follows the standard integrand reduction
	procedure, where the spurious monomials are present in the decomposed
	integrand, although they do not contribute to the integrated
	amplitude. Alternatively, if the dependence of the numerators on the
	loop momenta is known, then the integration over the orthogonal angles
	can be carried out before the reduction. Within this second option,
	which we can refer to as {\it integrate-divide}, after eliminating the
	orthogonal angles from the integrands, the residues resulting from the
	polynomial divisions contain only non-spurious monomials. In order to
	integrate before the reduction, the dependence of the numerator on the
	loop momenta should be either known analytically or reconstructed
	semi-analytically \cite{Heinrich:2010ax, Hirschi:2016mdz}. Such situation may indeed occur when
	the integrands to be reduced are built
	by automatic generators or they emerge as quotients of the subsequent
	divisions.

\section{Applications}
We summarise the results obtained from the application of the
\textit{adaptive} integrand decomposition (AID) at one loop in
Table~\ref{tab:1l}.
In the first column, $\boldsymbol{\tau}$ labels the variables the denominators depend on. $\Delta_{i_1\,\cdots\,i_n}$ indicates the residue obtained after the polynomial division of an arbitrary $n$-rank numerator and $\Delta_{i_1\,\cdots\,i_n}^{\text{int}}$ the result of its integral over transverse directions. $\Delta_{i_1\,\cdots\,i_n}^{'}$ corresponds to the minimal residue obtained from a further division of $\Delta_{i_1\,\cdots\,i_n}^{\text{int}}$. In the figures, wavy lines indicate massless particles, whereas solid ones stands for arbitrary masses.
	\begin{table}[h]
		\centering
		\renewcommand{\arraystretch}{1.2}
		\scalebox{0.77}{
			\begin{tabular}{|c c||c|c|c|c|}
				\hline
				\multicolumn{2}{|c||}{$\mathcal{I}_{i_1\,\cdots\,i_n}$}&$\boldsymbol{\tau}$&$\Delta_{i_1\,\cdots\,i_n}$ &$\Delta^{\text{int}}_{i_1\,\cdots\,i_n}$& $\Delta^{'}_{i_1\,\cdots\,i_n}$ \\
				\hline
				\hline
				\multirow{2}{0.9cm}{\centering $\mathcal{I}_{i_1i_2i_3i_4i_5}$}&\multirow{2}{2.cm}{\centering\includegraphics[height=0.37in]{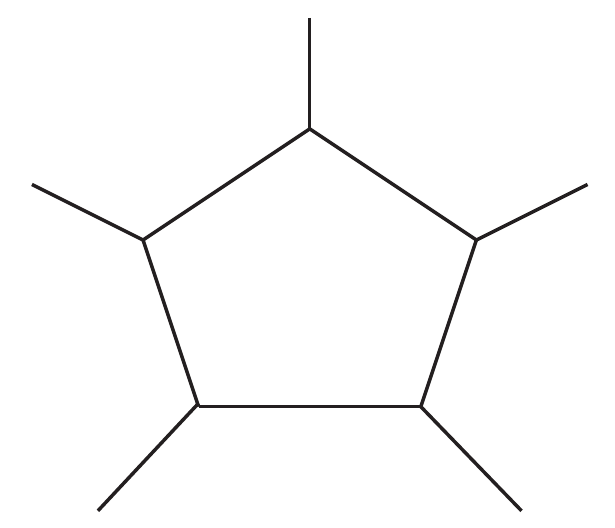}}&  &$1$&$-$&$-$\\
				&                &$\{x_1,x_2,x_3,x_4,\mu^2\}$ &$\{1\}$&$-$&$-$\\
				\hline
				\multirow{2}{0.9cm}{\centering $\mathcal{I}_{i_1i_2i_3i_4}$}&\multirow{2}{2.cm}{\centering\includegraphics[height=0.35in]{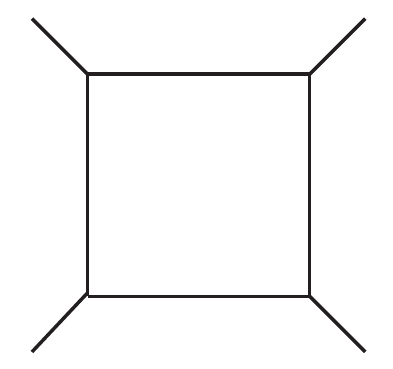}} & &$5$&$3$&$1$\\
				&     &$\{x_1,x_2,x_3,\lambda^2\}$ &$\{1,x_4,x_4^2,x_4^3,x_4^4\}$&$\{1,\lambda^{2},\lambda^4\}$&$\{1\}$\\
				\hline
				\multirow{2}{0.9cm}{\centering $\mathcal{I}_{i_1i_2i_3}$}&\multirow{2}{2.cm}{\centering\includegraphics[height=0.35in]{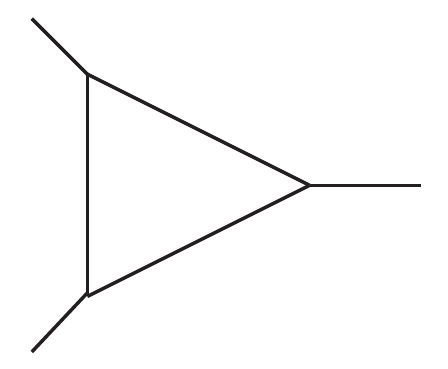}} &&$10$&$2$&$1$\\
				&    &  $\{x_1,x_2,\lambda^2\}$  &$\{1,x_3,x_4,x_3^2,x_3x_4,x_4^2,x_3^3,x_3^2x_4,x_3x_4^2,x_4^3\}$&$\{1,\lambda^{2}\}$&$\{1\}$\\
				\hline
				\multirow{2}{0.9cm}{\centering $\mathcal{I}_{i_1i_2}$}&\multirow{2}{2.cm}{\centering\includegraphics[height=0.35in]{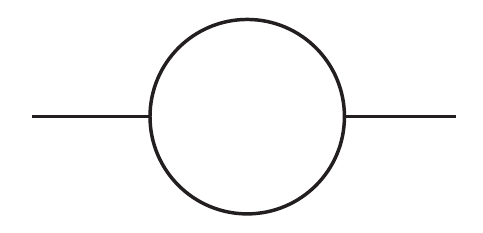}}&&$10$&$2$&$1$\\
				&        &$\{x_1,\lambda^2\}$&$\{1,x_2,x_3,x_4,x_2^2,x_2x_3,x_2x_4,x_3^2,x_3x_4,x_4^2\}$&$\{1,\lambda^2\}$&$\{1\}$\\
				\hline
				\multirow{2}{0.9cm}{\centering $\mathcal{I}_{i_1i_2}$}&\multirow{2}{2.cm}{\centering\includegraphics[height=0.35in]{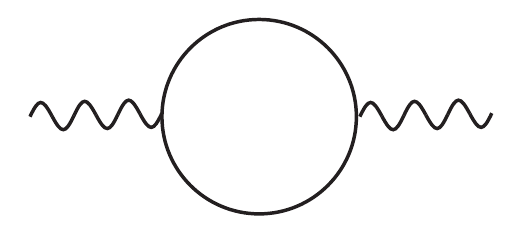}} &&$10$&$4$&$3$\\
				&     & $\{x_1,x_2,\lambda^2\}$  &$\{1,x_1,x_3,x_4,x_1^2,x_1x_3,x_1x_4,x_3^2,x_3x_4,x_4^2\}$&$\{1,x_1,x_1^2,\lambda^{2}\}$&$\{1,x_1,x_1^2\}$\\
				\hline
				\multirow{2}{0.9cm}{\centering $\mathcal{I}_{i_1}$}&\multirow{2}{2.cm}{\centering\includegraphics[height=0.35in]{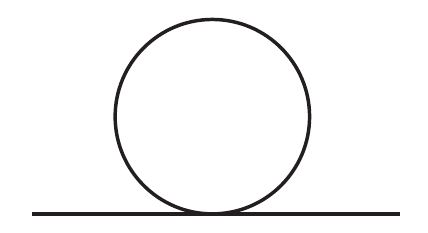}} &&$5$&$1$&$-$\\
				&    & $\{\lambda^2\}$   &$\{1,x_1,x_2,x_3,x_4\}$&$\{1\}$&$-$\\
				\hline
			\end{tabular}
		}
		\caption{\small{Residue parametrisation for irreducible one-loop topologies.}}
		\label{tab:1l}
	\end{table}
As an exceptional property of one-loop integrands, we find that by working with $\boldsymbol{\tau}$ variables, all unitarity cuts are reduced to zero-dimensional systems. Moreover, we show that the last step of the algorithm, \textit{i.e.} the further polynomial division after angular integration over the transverse space, provides an implementation of the dimensional recurrence relations at the integrand level. 

\begin{figure*}[ht!]
	 	\centering
	 	\begin{subfigure}[t]{0.33\textwidth}
	 		\centering
	 		\includegraphics[height=0.8in]{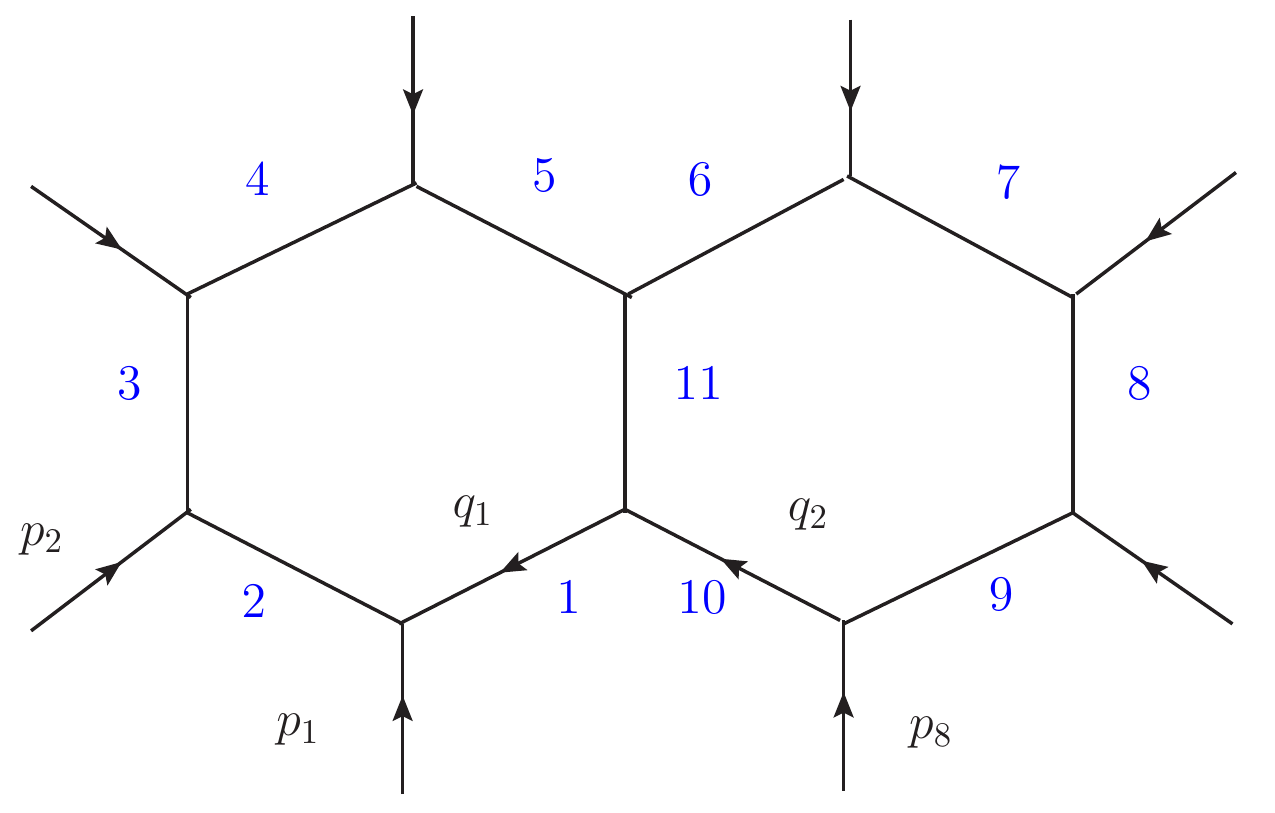}
	 		\caption{$\mathcal{I}_{12345678910\, 11}^{\text{P}}$}
	 		\label{fig:maxcutP}
	 	\end{subfigure}%
	 	\begin{subfigure}[t]{0.33\textwidth}
	 		\centering
	 		\includegraphics[height=0.9in]{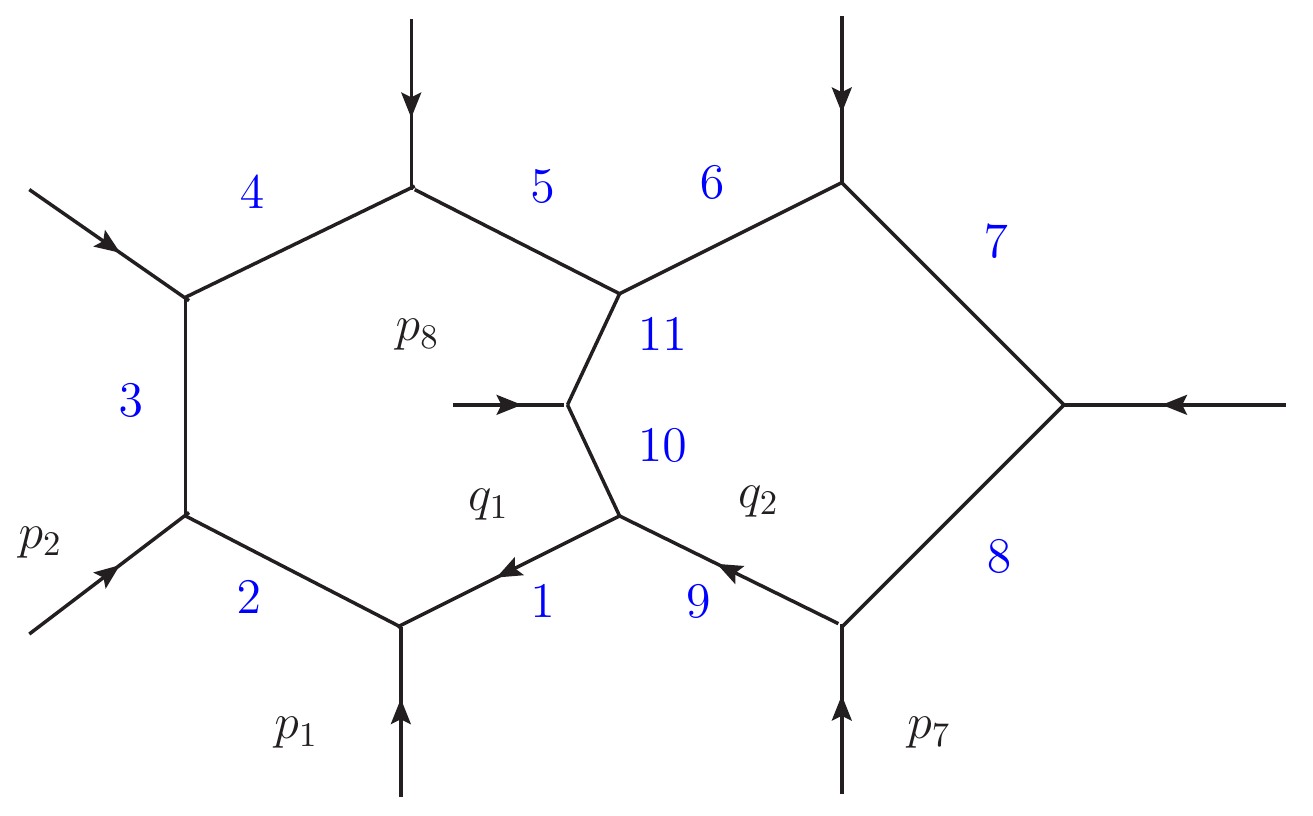}
	 		\caption{$\mathcal{I}_{12345678910\, 11}^{\text{NP}1}$}
	 		\label{fig:maxcutNP1} 
	 	\end{subfigure}
	 	\begin{subfigure}[t]{0.33\textwidth}
	 		\centering
	 		\includegraphics[height=0.9in]{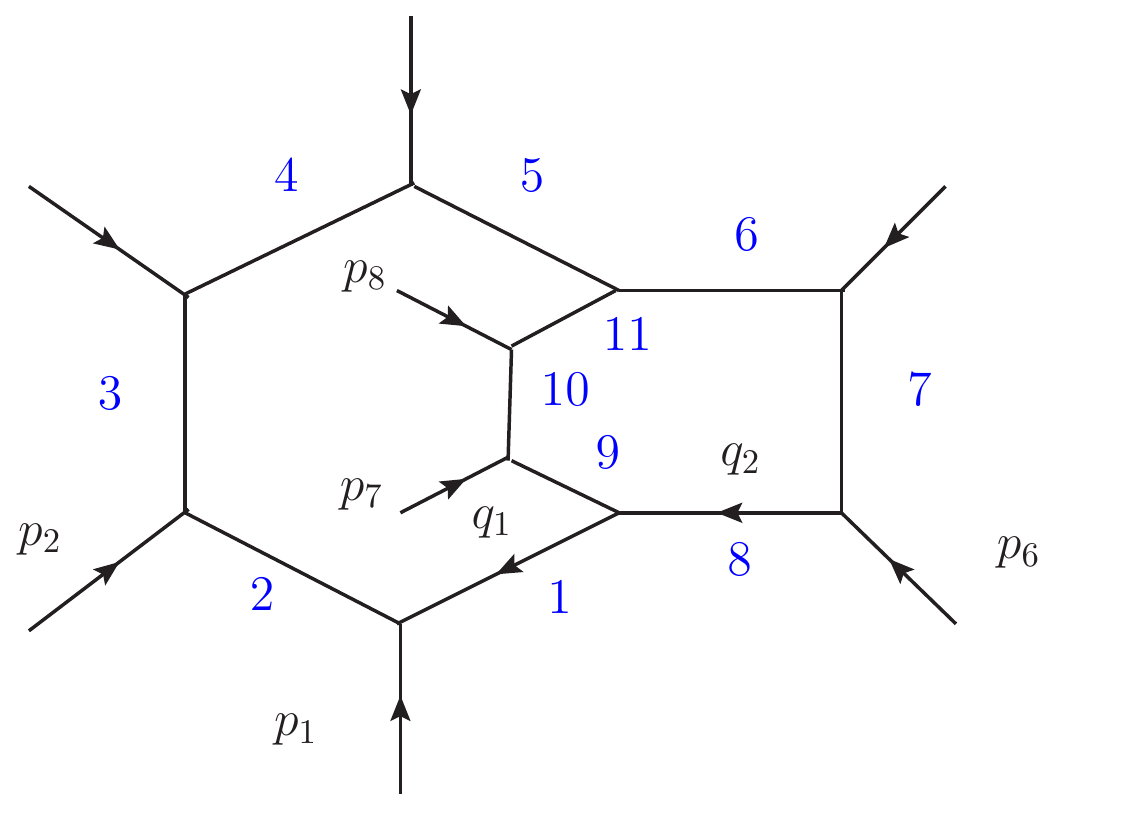}
	 		\caption{$\mathcal{I}_{12345678910\, 11}^{\text{NP2}}$}
	 		\label{fig:maxcutNP2}
	 	\end{subfigure}
	 	\caption{Maximum-cut topologies}
	 	\label{fig:maxcut}
	 \end{figure*}

Beside revisiting the one-loop, we applied the AID in order to determine
the universal parametrisation of the residues appearing in the
integrand decomposition \eqref{eq:intdec} of the three eight-point
topologies shown in fig.~\ref{fig:maxcutP}-\ref{fig:maxcutNP2}. The
results obtained are valid for arbitrary (internal and external)
kinematic configuration.
For the complete results of the two-loop decomposition, we refer the
reader to the ref. \cite{Mastrolia:2016dhn}.

\begin{figure}
\centering
\includegraphics[scale=0.85]{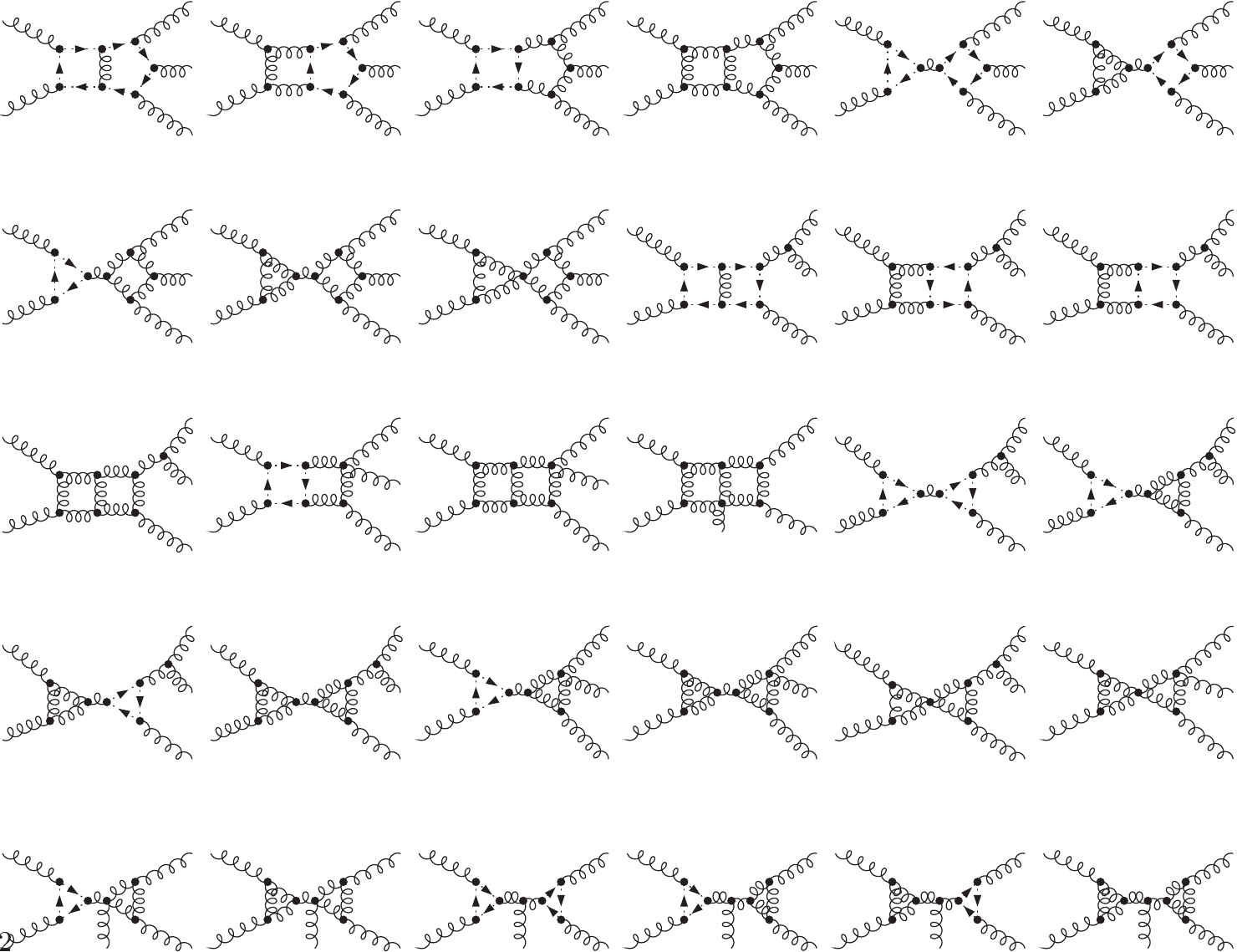}
\caption{Selection of Feynman diagrams contributing to the five-gluons amplitude. Curly lines represent gluons and dashed ones indicate ghosts.}
\label{fig:5ptdiag}
\end{figure}

Furthermore, we applied the AID to the leading color contribution to
the two-loop all-plus five-gluon amplitude \cite{Badger:2013gxa,Badger:2015lda,Papadopoulos:2015jft,Gehrmann:2015bfy,Dunbar:2016aux,Dunbar:2016cxp,Badger:2016ozq}, which, after the first step of the division algorithm, admits a decomposition of the form
\begin{multline}
A^{(2)}(1^{+},2^{+},3^{+},4^{+},5^{+})=
\int \frac{d^{d}q_{1}}{\pi^{d/2}}\frac{d^{d}q_{2}}{\pi^{d/2}} \ \Bigg\{
\frac{\Delta\bigg(\parbox[h][0.07\linewidth][c]{0.09\linewidth}{
\centering
  \includegraphics[scale=0.5]{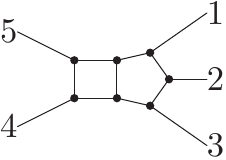}}\bigg)}{D_{1}\,D_{2}\,D_{3}\,D_{4}\,D_{5}\,D_{6}\,D_{7}\,D_{8}}
+\frac{\Delta\bigg(\parbox[h][0.07\linewidth][c]{0.09\linewidth}{
\centering
  \includegraphics[scale=0.5]{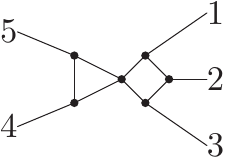}}\bigg)}{D_{1}\,D_{2}\,D_{3}\,D_{4}\,D_{5}\,D_{6}\,D_{7}}
  \\+\frac{\Delta\bigg(\parbox[h][0.07\linewidth][c]{0.09\linewidth}{
\centering
  \includegraphics[scale=0.5]{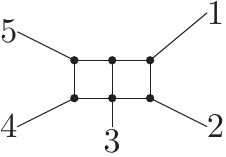}}\bigg)}
  {D_{1}\,D_{2}\,D_{3}\,D_{5}\,D_{6}\,D_{7}\,D_{8}}
+\frac{\Delta\bigg(
\parbox[h][0.07\linewidth][c]{0.09\linewidth}{
\centering
  \includegraphics[scale=0.5]{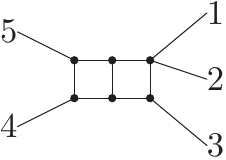}}\bigg)}{D_{1}\,D_{3}\,D_{4}\,D_{5}\,D_{6}\,D_{7}\,D_{8}}
  +\frac{\Delta\bigg(
\parbox[h][0.07\linewidth][c]{0.09\linewidth}{
\centering
  \includegraphics[scale=0.5]{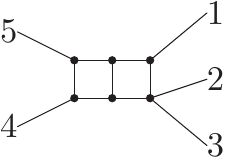}}\bigg)}
  {D_{1}\,D_{2}\,D_{4}\,D_{5}\,D_{6}\,D_{7}D_{8}}\\
  +\frac{\Delta\bigg(\parbox[h][0.07\linewidth][c]{0.09\linewidth}{
\centering
  \includegraphics[scale=0.5]{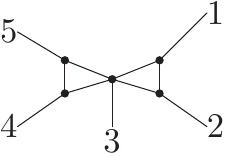}}\bigg)}{D_{1}\,D_{2}\,D_{3}\,D_{5}\,D_{6}\,D_{7}}
  +\frac{\Delta\bigg(\parbox[h][0.07\linewidth][c]{0.09\linewidth}{
\centering
 \includegraphics[scale=0.5]{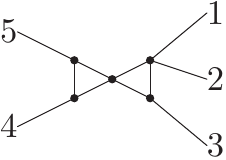}}\bigg)}{D_{1}\,D_{3}\,D_{4}\,D_{5}\,D_{6}\,D_{7}}
 +\frac{\Delta\bigg(\parbox[h][0.07\linewidth][c]{0.09\linewidth}{
\centering
  \includegraphics[scale=0.5]{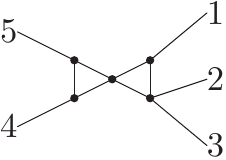}}\bigg)}{D_{1}\,D_{2}\,D_{4}\,D_{5}\,D_{6}\,D_{7}}
\Bigg\}\\
+\text{cycl. perm.}
\end{multline}
The residue have been obtained from numerators constructed through Feynman diagrams in Feynman gauge, including both gluon and ghost loop contributions. The relevant Feynman graphs, a selection of which is shown in fig.~\ref{fig:5ptdiag}, have been generated by using \Feynarts~\cite{Hahn:2000kx} and \Feyncalc~\cite{Mertig:1990an, Shtabovenko:2016sxi}. The expression of the residues have been numerically checked against the results of~\cite{Badger:2013gxa}. 
The integration of the transverse directions of both four-point and factorised topologies and the further division of the integrated residues may lead to a new representation of the amplitude, whose discussion is, nevertheless,  beyond the scope of this report.
\section{Conclusions}

Owing to the representation of Feynman integrals in
parallel and orthogonal space, numerators and denominators of integrands
appear to depend on different sets of integration variables.
By exploiting the different origin and role of these variables,
we engineered a novel variant of the integrand decomposition algorithm, 
defined as {\it adaptive integrand decomposition} (AID),
where the multivariate polynomial division is simplified and the integration over
transverse space variables can be efficiently carried out by means of Gegenbauer polynomials.
Orthogonality relations for Gegenbauer polynomials detect and
annihilate the so called spurious integrals at any loop order, and
can be used any time that a certain subset of integration
variables do not appear in the denominators, as it happens also in the
case of factorised diagrams and ladder topologies. 

As a result of the AID, each amplitude is written in terms of a
set of integrals which, beside the scalar ones, contains tensor
integrals with irreducible scalar products depending on the 
parallel directions and on the lengths of the transverse vectors only.
In addition, we have shown that the integration over the transverse directions leads to integrals which
can be subject to additional polynomial divisions, which in some cases
correspond to dimension-shifting recurrence relations implemented at the integrand level.

We revisited the one-loop integrand decomposition, showing 
that it is completely determined by the maximum-cut theorem in
different dimensions. Furthermore,
we considered the complete reduction of two-loop planar and
non-planar integrals for arbitrary kinematics, classifying the
corresponding residues and identifying the independent integrals
contributing to eight-particle scattering amplitudes. 
The proposed algorithm can be simply extended to higher loops.

In view of  the development of automated tools for the evaluation
of multi-loop amplitudes,
the AID can be used as an intermediate reduction phase to achieve an
expression in terms of an
independent set of integrals. The latter can be further simplified by means of
relations arising from additional symmetries that can
occur, such as integration-by-parts identities and color-kinematics duality.
%

\bibliographystyle{JHEP-LL12}
\bibliography{references}

\providecommand{\href}[2]{#2}\begingroup\raggedright\begin{thebibliography}{10}

\bibitem{Ossola:2006us}
G.~Ossola, C.~G. Papadopoulos, and R.~Pittau {\em Nucl. Phys.} {\bf B763}
  (2007) 147--169, [\href{http://xxx.lanl.gov/abs/hep-ph/0609007}{{\tt
  hep-ph/0609007}}].

\bibitem{Ossola:2007bb}
G.~Ossola, C.~G. Papadopoulos, and R.~Pittau {\em JHEP} {\bf 07} (2007) 085,
  [\href{http://xxx.lanl.gov/abs/0704.1271}{{\tt 0704.1271}}].

\bibitem{Ellis:2007br}
R.~K. Ellis, W.~T. Giele, and Z.~Kunszt {\em JHEP} {\bf 03} (2008) 003,
  [\href{http://xxx.lanl.gov/abs/0708.2398}{{\tt 0708.2398}}].

\bibitem{Ellis:2008ir}
R.~K. Ellis, W.~T. Giele, Z.~Kunszt, and K.~Melnikov {\em Nucl. Phys.} {\bf
  B822} (2009) 270--282, [\href{http://xxx.lanl.gov/abs/0806.3467}{{\tt
  0806.3467}}].

\bibitem{Ossola:2008xq}
G.~Ossola, C.~G. Papadopoulos, and R.~Pittau {\em JHEP} {\bf 05} (2008) 004,
  [\href{http://xxx.lanl.gov/abs/0802.1876}{{\tt 0802.1876}}].

\bibitem{Mastrolia:2008jb}
P.~Mastrolia, G.~Ossola, C.~G. Papadopoulos, and R.~Pittau {\em JHEP} {\bf 06}
  (2008) 030, [\href{http://xxx.lanl.gov/abs/0803.3964}{{\tt 0803.3964}}].

\bibitem{Mastrolia:2012bu}
P.~Mastrolia, E.~Mirabella, and T.~Peraro {\em JHEP} {\bf 06} (2012) 095,
  [\href{http://xxx.lanl.gov/abs/1203.0291}{{\tt 1203.0291}}]. [Erratum:
  JHEP11,128(2012)].

\bibitem{Ossolaproc}
G.~Ossola {\em These proceedings}.

\bibitem{Mastrolia:2011pr}
P.~Mastrolia and G.~Ossola {\em JHEP} {\bf 11} (2011) 014,
  [\href{http://xxx.lanl.gov/abs/1107.6041}{{\tt 1107.6041}}].

\bibitem{Badger:2012dp}
S.~Badger, H.~Frellesvig, and Y.~Zhang {\em JHEP} {\bf 04} (2012) 055,
  [\href{http://xxx.lanl.gov/abs/1202.2019}{{\tt 1202.2019}}].

\bibitem{Zhang:2012ce}
Y.~Zhang {\em JHEP} {\bf 09} (2012) 042,
  [\href{http://xxx.lanl.gov/abs/1205.5707}{{\tt 1205.5707}}].

\bibitem{Mastrolia:2012an}
P.~Mastrolia, E.~Mirabella, G.~Ossola, and T.~Peraro {\em Phys. Lett.} {\bf
  B718} (2012) 173--177, [\href{http://xxx.lanl.gov/abs/1205.7087}{{\tt
  1205.7087}}].

\bibitem{Ita:2015tya}
H.~Ita \href{http://xxx.lanl.gov/abs/1510.05626}{{\tt 1510.05626}}.

\bibitem{Larsen:2015ped}
K.~J. Larsen and Y.~Zhang {\em Phys. Rev.} {\bf D93} (2016), no.~4 041701,
  [\href{http://xxx.lanl.gov/abs/1511.01071}{{\tt 1511.01071}}].

\bibitem{VonManteuffel:2014ixa}
A.~von Manteuffel and R.~M. Schabinger {\em Phys. Lett.} {\bf B744} (2015)
  101--104, [\href{http://xxx.lanl.gov/abs/1406.4513}{{\tt 1406.4513}}].

\bibitem{Kant:2013vta}
P.~Kant {\em Comput. Phys. Commun.} {\bf 185} (2014) 1473--1476,
  [\href{http://xxx.lanl.gov/abs/1309.7287}{{\tt 1309.7287}}].

\bibitem{Henn:2013pwa}
J.~M. Henn {\em Phys. Rev. Lett.} {\bf 110} (2013) 251601,
  [\href{http://xxx.lanl.gov/abs/1304.1806}{{\tt 1304.1806}}].

\bibitem{Argeri:2014qva}
M.~Argeri, S.~Di~Vita, P.~Mastrolia, E.~Mirabella, J.~Schlenk, U.~Schubert, and
  L.~Tancredi {\em JHEP} {\bf 03} (2014) 082,
  [\href{http://xxx.lanl.gov/abs/1401.2979}{{\tt 1401.2979}}].

\bibitem{Papadopoulos:2014lla}
C.~G. Papadopoulos {\em JHEP} {\bf 07} (2014) 088,
  [\href{http://xxx.lanl.gov/abs/1401.6057}{{\tt 1401.6057}}].

\bibitem{Borowka:2015mxa}
S.~Borowka, G.~Heinrich, S.~P. Jones, M.~Kerner, J.~Schlenk, and T.~Zirke {\em
  Comput. Phys. Commun.} {\bf 196} (2015) 470--491,
  [\href{http://xxx.lanl.gov/abs/1502.06595}{{\tt 1502.06595}}].

\bibitem{Smirnov:2015mct}
A.~V. Smirnov {\em Comput. Phys. Commun.} {\bf 204} (2016) 189--199,
  [\href{http://xxx.lanl.gov/abs/1511.03614}{{\tt 1511.03614}}].

\bibitem{Mastrolia:2016dhn}
P.~Mastrolia, T.~Peraro, and A.~Primo
  \href{http://xxx.lanl.gov/abs/1605.03157}{{\tt 1605.03157}}.

\bibitem{Collins:105730}
J.~C. Collins, {\em {Renormalization: an introduction to renormalization, the
  renormalization group, and the operator-product expansion}}.
\newblock Cambridge monographs on mathematical physics. Cambridge Univ. Press,
  Cambridge, 1984.

\bibitem{Kreimer:1991wj}
D.~Kreimer {\em Z. Phys.} {\bf C54} (1992) 667--672.

\bibitem{Kreimer:1992zv}
D.~Kreimer {\em Phys. Lett.} {\bf B292} (1992) 341--347.

\bibitem{Czarnecki:1994td}
A.~Czarnecki, U.~Kilian, and D.~Kreimer {\em Nucl. Phys.} {\bf B433} (1995)
  259--275, [\href{http://xxx.lanl.gov/abs/hep-ph/9405423}{{\tt
  hep-ph/9405423}}].

\bibitem{Frink:1996ya}
A.~Frink, U.~Kilian, and D.~Kreimer {\em Nucl. Phys.} {\bf B488} (1997)
  426--440, [\href{http://xxx.lanl.gov/abs/hep-ph/9610285}{{\tt
  hep-ph/9610285}}].

\bibitem{kreimer:1996qy}
D.~Kreimer {\em Nucl. Instrum. Meth.} {\bf A389} (1997) 323--326.

\bibitem{Mastrolia:2013kca}
P.~Mastrolia, E.~Mirabella, G.~Ossola, and T.~Peraro {\em Phys. Lett.} {\bf
  B727} (2013) 532--535, [\href{http://xxx.lanl.gov/abs/1307.5832}{{\tt
  1307.5832}}].

\bibitem{Heinrich:2010ax}
G.~Heinrich, G.~Ossola, T.~Reiter, and F.~Tramontano {\em JHEP} {\bf 10} (2010)
  105, [\href{http://xxx.lanl.gov/abs/1008.2441}{{\tt 1008.2441}}].

\bibitem{Hirschi:2016mdz}
V.~Hirschi and T.~Peraro \href{http://xxx.lanl.gov/abs/1604.01363}{{\tt
  1604.01363}}.

\bibitem{Badger:2013gxa}
S.~Badger, H.~Frellesvig, and Y.~Zhang {\em JHEP} {\bf 12} (2013) 045,
  [\href{http://xxx.lanl.gov/abs/1310.1051}{{\tt 1310.1051}}].

\bibitem{Badger:2015lda}
S.~Badger, G.~Mogull, A.~Ochirov, and D.~O'Connell {\em JHEP} {\bf 10} (2015)
  064, [\href{http://xxx.lanl.gov/abs/1507.08797}{{\tt 1507.08797}}].

\bibitem{Papadopoulos:2015jft}
C.~G. Papadopoulos, D.~Tommasini, and C.~Wever {\em JHEP} {\bf 04} (2016) 078,
  [\href{http://xxx.lanl.gov/abs/1511.09404}{{\tt 1511.09404}}].

\bibitem{Gehrmann:2015bfy}
T.~Gehrmann, J.~M. Henn, and N.~A. Lo~Presti {\em Phys. Rev. Lett.} {\bf 116}
  (2016), no.~6 062001, [\href{http://xxx.lanl.gov/abs/1511.05409}{{\tt
  1511.05409}}]. [Erratum: Phys. Rev. Lett.116,no.18,189903(2016)].

\bibitem{Dunbar:2016aux}
D.~C. Dunbar and W.~B. Perkins {\em Phys. Rev.} {\bf D93} (2016), no.~8 085029,
  [\href{http://xxx.lanl.gov/abs/1603.07514}{{\tt 1603.07514}}].

\bibitem{Dunbar:2016cxp}
D.~C. Dunbar, G.~R. Jehu, and W.~B. Perkins {\em Phys. Rev.} {\bf D93} (2016),
  no.~12 125006, [\href{http://xxx.lanl.gov/abs/1604.06631}{{\tt 1604.06631}}].

\bibitem{Badger:2016ozq}
S.~Badger, G.~Mogull, and T.~Peraro
  \href{http://xxx.lanl.gov/abs/1606.02244}{{\tt 1606.02244}}.

\bibitem{Hahn:2000kx}
T.~Hahn {\em Comput. Phys. Commun.} {\bf 140} (2001) 418--431,
  [\href{http://xxx.lanl.gov/abs/hep-ph/0012260}{{\tt hep-ph/0012260}}].

\bibitem{Mertig:1990an}
R.~Mertig, M.~Bohm, and A.~Denner {\em Comput. Phys. Commun.} {\bf 64} (1991)
  345--359.

\bibitem{Shtabovenko:2016sxi}
V.~Shtabovenko, R.~Mertig, and F.~Orellana
  \href{http://xxx.lanl.gov/abs/1601.01167}{{\tt 1601.01167}}.

\end{thebibliography}\endgroup

\end{document}